# INITIAL HIGH ELECTRIC FIELD – VACUUM ARC BREAKDOWN TEST RESULTS FOR ADDITIVELY MANUFACTURED PURE COPPER ELECTRODES*

A. Ratkus†, T. Torims, G. Pikurs, Riga Technical University, Riga, Latvia
V. Bjelland, S. Calatroni, R. Peacock, C. Serafim, M. Vretenar, W. Wuensch
CERN, Geneva, Switzerland
M. Vedani, T. Romano[1], Politecnico di Milano, Milan, Italy
M. Pozzi, M. Foppa Pedretti, Rösler Italiana s.r.l., Milan, Italy
[1] also at Riga Technical University, Riga, Latvia

## Abstract

Additive Manufacturing (AM) is already well-established for various manufacturing applications, providing many benefits such as design freedom, novel and complex cooling designs for the parts and different performance improvements, as well as significantly reducing the production time. With the mentioned characteristics, AM is also being considered as a technology for manufacturing a Radio Frequency Quadrupole (RFQ) prototype. For this application, an important parameter is the voltage holding capability of the surfaces. Furthermore, the voltage holding capability of pure copper surfaces manufactured by AM is of interest for the accelerator community at large for prospective future developments. To characterize these properties, a series of high electric field tests were performed on pure copper electrodes produced by AM, using the CERN pulsed high-voltage DC system. The tests were carried out with AM produced electrodes with large surface roughness. During the testing process, a high vacuum was maintained. The electric breakdown rate was also monitored to ensure not to exceed the breakdown limit of $10^{-5}$ breakdowns per pulse. The achieved results provide the first, initial reference values for the performance of AM built pure copper electrodes for vacuum arc breakdown testing. Initial results prove the capability of AM electrodes to hold a high electric field, while having low breakdown rates. These are crucial results for further AM technology usage for different AM pure-copper accelerator components.

## INTRODUCTION

AM technology has found more and more applications for various accelerator components due to the design freedom, while maintaining similar characteristics for AM pure copper compared to bulk copper [1]. Furthermore, AM technology has shown to be a valid candidate for the production of complex accelerator structures, i.e. compact Radio Frequency Quadrupoles (RFQs) in the very near future [2]. However, for further developments AM produced component surfaces face many challenges, inter alia. Highest achievable vacuum levels and electrical field gradient can be limited by the material properties, surface conditions, roughness, defects and inclusions.

Vacuum arc breakdowns are a limiting factor in high electric field applications. It is therefore crucial to find design and manufacturing methods to minimise the probability of breakdowns.

Existing data typically focus on smooth test surfaces [3], therefore, information about AM components with large surface roughness is missing, to assess the performance in high electrical field applications. For that reason, high voltage (HV) holding tests were performed to acquire the first reference values for the AM built, pure copper electrode performance for high electrical field applications.

## MATERIALS, EQUIPMENT AND METHODS

### The High Electrical Field Test and Setup

A pulsed high-voltage DC system available at CERN, depicted in Fig.1 a), was used for the high electrical field tests. The applied voltage is pulsed at a frequency of 1000 Hz and has a pulse length of 1 µs.

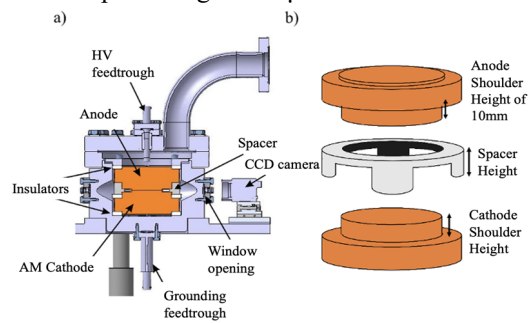

Figure 1: a) CERN's pulsed high-voltage DC system. b) Anode, cathode and insulating spacer. [3, 4].

The system consists of a vacuum chamber containing two custom-made electrodes, an anode and a cathode as seen in Fig. 1b). An insulating spacer is used for separating the electrodes, creating a custom gap height. This allows to control the highest electric field reachable in the system, as the maximum applied voltage is 7 kV. A more detailed description of the test methodology and the experimental system is given elsewhere [3, 4].

* This project has received funding from the European Union's Horizon 2020 Research and Innovation programme under grant agreement No 101004730 and is supported by the Latvian Council of Science under grant agreement VPP-IZM-CERN-2022/1-0001.
† andris.ratkus@rtu.lv

### The Test Electrodes

It has been established, that vacuum arcs originate from the cathode. Therefore, a cathode was produced with AM without any additional machining on the area exposed to the high electric field. The anode used for testing was produced with conventional manufacturing methods.

The anode was machined from a pure, bulk copper workpiece to ensure a $R_a$ = 0.4 μm on the effective test surface having a shoulder height of 10 mm. The cathode, on the other hand, was built with EOS M280 LPBF system by using 99.95 % pure Cu powder, with particle size distribution parameters as follows: D10 = 21.1 μm, D50 = 27.4 μm, D90 = 37.6 μm. The shoulder height (SH) of the cathode was machined to manipulate the gap size of the system, as the spacers have static heights.

### AM Cathode Characterization

The AM cathode was tested in a as-built condition, with additional machining only on the support shoulder for the spacer. Surface roughness measurements were performed using the Mitutoyo contact profilometer Formtracer Avant FTA equipment (ISO 4288:1997). Surface roughness values for two perpendicular measurements were found to be $R_a$ = 8.28 ± 0.89 and 10.67 ± 1.16 μm, with $R_z$ = 42.10 ± 5.33 and 52.76 ± 7.56 μm, respectively. Additional 3D scans using GOM Atos Compact Scan are shown in Fig.2.

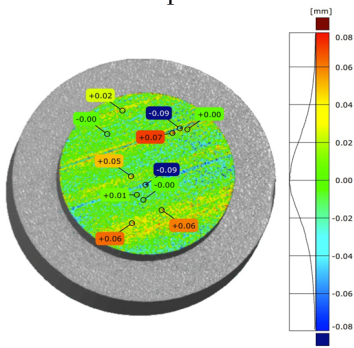

Figure 2: AM Cathode 3D scan.

### The Gap Height Control and Reduction

The gap height between the electrodes determines the highest reachable field. Due to the relatively high surface roughness of the AM electrode, the gap size was reduced in steps by increasing the shoulder height (SH) (see Fig. 1 b)). While varying the gap height, electrode mechanical measurements were done by using TRIMOS V9 vertical measurement equipment and ZEISS Prismo Ultra 12-18-10 (CMM) (1.2 μm +L/500 mm; F= 0.05 N; v= 5 mm/min).

A SH of 10 mm for the anode was ensured during manufacturing and was not changed during the testing process. AM cathode's SH was machined for adjustment of the gap height between the electrodes (Fig. 3). For the first test, a gap size of 270 μm was used having a SH of 9.78 mm and a spacer of 20.02 mm. For the second test, a gap size of 115 μm was used having a SH of 9.975 mm and a spacer of 20.06 mm.

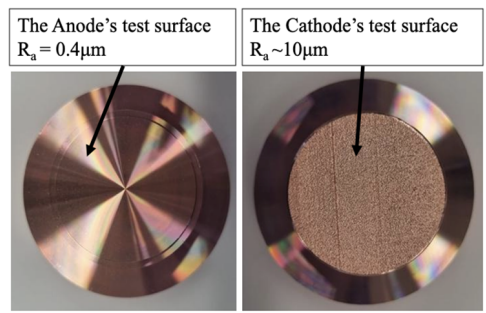

Figure 3: Conventionally manufactured anode and AM built cathode with the machined shoulders.

## RESULTS

HV tests were conducted by using the same electrodes with gap heights between the electrodes of 270 μm (Test 1) and 115 μm (Test 2), therefore Test 2 starts with the already electrode conditioned. The results obtained show the vacuum and high electric field holding parameters of the AM manufactured cathode.

### The Vacuum Pump Down

Figures 4 a) and b) show the initial vacuum pump down for Tests 1 and 2 after the standard vacuum cleaning process, respectively.

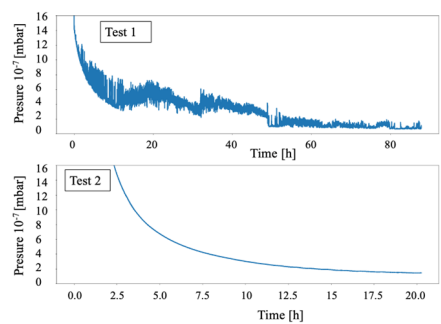

Figure 4: Vacuum pump down for Tests 1 and 2.

For Test 1 approximately 55 hours are needed to reach a stable vacuum with minor fluctuations. In case of Test 2 only 3 hours are needed to reach an acceptable and stable level of vacuum. It is observed that the pressure fluctuations are much lower for the case of Test 2. It is possible to reach a high vacuum level, while using AM electrodes. The final vacuum level for Test 1 was $3 \cdot 10^{-8}$ mbar, while for Test 2 – $2 \cdot 10^{-8}$ mbar.

### High Voltage Testing

Figure 5 shows the HV testing results for Tests 1 and 2. Test 1 reached a stable, maximum electric field of 26 MV/m, equivalent to the system maximum voltage of 7 kV. Under the same conditions, for Test 2 an electric field of 40 MV/m is reached but has yet to reach the maximum value. The maximum accepted breakdown rate is at $1 \cdot 10^{-5}$ breakdowns/pulse. The breakdown rates achieved with the AM electrodes were far below this value and were stable for the duration of the experiment. It should be noted that the achieved electric field gradient of 40 MV/m for Test 2 corresponds to the operating conditions of the compact 750 MHz RFQ design of CERN [5].

Breakdown locations were observed for Test 1, while no specific locations were observed for Test 2. This is due to light scattering and/or absorption caused by the high surface roughness of the cathode, therefore limiting the light signal reaching both of the cameras.

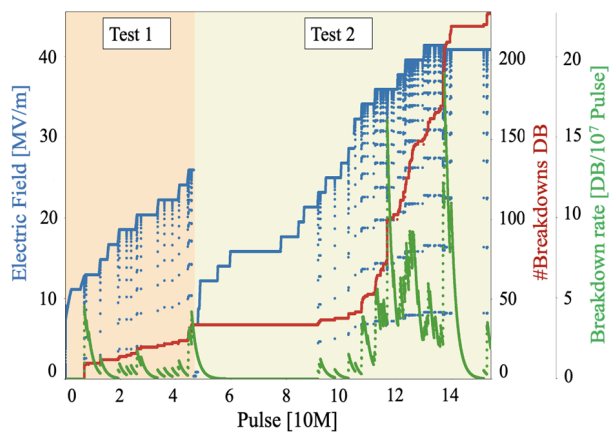

Figure 5: Electrode conditioning of the AM electrode.

## DISCUSSION

For 270 μm gap height (Test 1), the first breakdown is observed with an electric field of 12 MV/m. As a reference, for an oxygen-free and heat-treated electrode with low surface roughness measured with a gap height of 60 μm [6], the first breakdowns are observed at a field of approximately 5 MV/m. It is also observed that the AM electrodes experience less breakdowns before reaching an electricfield of 25 MV/m, yielding a lower breakdown rate. At the same time, it can noted that the AM electrodes needed approximately twice as many pulses to reach the equivalent field, compared to the reference case.

For 115 μm gap (Test 2), the first breakdowns do not occur until an electric field of 23 MV/m is reached. It can noted that the electrodes have already undergone conditioning to this field level and therefore not unexpected to hold a higher field than for 270 μm gap height. It can be seen in the conditioning data in Fig. 5 that a low breakdown rate is still achieved even after exceeding 30 MV/m. The electrode seems to have had a clustering of breakdowns around 120M and 140M pulses, which is not uncommon based on previous experience. Without the possibility of localizing the position of breakdowns, it is difficult to say if they all occurred at the same spot on the electrode or were spread over the surface. In the latter case, the applied field was slightly reduced to stabilise the system. Although the electrode has not completed conditioning, it is close to its final electric field value.

The surface roughness and chosen material characteristics usually are the biggest limiting factors to the highest achievable electric field [6]. Thus, the surface roughness of the AM electrodes and usage of pure copper could be the limiting factors in the case of these tests. Therefore, HV tests with AM cathodes, with a reduced gap height should be continued to better grasp the limitations of AM electrodes. Since high mechanical tolerances are difficult to ensure with a large surface roughness, physical contact between the two electrodes could be the limiting factor for the minimum achievable gap.

## CONCLUSION

Initial tests show the capability of AM electrodes to hold high vacuum levels and high electric fields with low breakdown rates, approving AM to be a valid candidate for accelerator component manufacturing. The electrodes were tested with two gap heights – 270 μm and 115 μm. The number of breakdowns and breakdown rate are found to be even lower than for a reference of a standard, oxygen-free, heat-treated copper electrodes. During the tests, slower conditioning was applied compared to the reference, using approximately twice as many pulses. A stable, electric field of 40 MV/m was reached with the smaller gap height of 115 μm.

Further tests with even smaller gap height should be continued to explore more details of the performance of AM component surfaces in high electric field applications. As the next step, the tests should be repeated with an AM electrode produced in different built directions for comparison of results and better understanding of the AM limitations. Furthermore, more tests should be performed of AM cathodes after surface finishing processes, using different state-of-the-art technologies.